\documentclass[aps,prd,twocolumn,nofootinbib]{revtex4}
\usepackage{graphicx}

\input epsf.sty

\usepackage{amsmath,amssymb,epsfig,color, wasysym, bm}

\newcommand{\hn}{\hat{\mathbf{n}}}
\newcommand{\be}{\begin{equation}}
\newcommand{\ee}{\end{equation}}
\newcommand{\ben}{\begin{eqnarray}}
\newcommand{\een}{\end{eqnarray}}
\newcommand{\lcdm}{\Lambda\text{CDM}}
\newcommand{\lb}{\left(}
\newcommand{\rb}{\right)}

\begin{document}

%
%
%
%
%

\title{CMB Lensing and Giant Rings}
\author{Ben Rathaus}
\email{ben.rathaus@gmail.com}
\author{Nissan Itzhaki}
\email{nitzhaki@post.tau.ac.il}

\affiliation{Raymond and Beverly Sackler Faculty of Exact Sciences,
School of Physics and Astronomy, Tel-Aviv University, Ramat-Aviv, 69978,
Israel}
\begin{abstract}
We study the CMB lensing signature of a pre-inationary particle (PIP), assuming it is responsible for the giant rings anomaly that was found recently in the WMAP data. Simulating Planck-like data we find that
generically the CMB lensing signal to noise ratio associated with such a PIP is quite small and it would be difficult to cross correlate the temperature giant rings with the CMB lensing signal. However, if the pre-inationary particle is also responsible for the bulk flow measured from the local large scale structure, which happens to point roughly at the same direction as the giant rings,  then the CMB lensing signal to noise ratio is fairly significant.

\end{abstract}
\maketitle

Recently, a new  anomaly was added to the list of large scale anomalies (for a review of large scale anomalies see \cite{Copi:2010na, WMAP7} and references within) - giant rings in the CMB \cite{Kovetz:2010kv}. While the statistical significance of this novel anomaly is moderate, about $3\sigma$, it is fascinating for two reasons. First, such rings are a distinct imprint of the model of a pre-inflationary particle (PIP),
 \cite{Kovetz:2010kv, Fialkov:2009xm} (it is this model that motivated the search for the giant rings). 
Second, the rings appear to be correlated with another large scale anomaly - the large bulk flow inferred from the large scale structure in our vicinity, reported in \cite{Sarkar:2006gh, Feldman:2008rm, arXiv:0809.4041,arXiv:0911.5516, arXiv:1111.0631}. Within the error of measurement they  point at the same direction, which according to \cite{Fialkov:2009xm} is the origin of the giant structure seeded by the PIP.

In this note we assume that a PIP is indeed responsible for the giant rings and study its  CMB  lensing signal. 
We start with a brief description of the PIP. As shown in \cite{Itzhaki:2008ih, Fialkov:2009xm} the effect of a PIP is to generate the following gravitational potential
\be
\Phi_{\text{PIP}}(k)=\left.\frac{\lambda H}{12 \sqrt{\pi \epsilon}
k^3}\right|_{k=a(t) H},
\ee
where $\epsilon$ is the slow roll parameter, $\lambda= dm/d\phi
-m\sqrt{\epsilon/2} $ (with $\phi$ being the inflaton field and $m$ the mass of the PIP) and, as
usual, the potential is evaluated at horizon crossing.

For constant $\lambda$ (and $n_s=1$) we find (using COBE normalization) a simple potential as a function of the distance from the PIP 
\be\label{eq:log}
\Phi_{\text{PIP}}(r,z=0)=\lambda C \log(r), 
 \ee
where
\be 
C=1.09 \times 10^{-5},
\ee
which is easy to work with.  There is, however, no reason why  $\lambda$ should be  constant. Theoretically we know that in the stringy model of inflation that initiated this investigation \cite{Itzhaki:2007nk} $\lambda$ is not a constant.  In addition, experimentally we know that (\ref{eq:log}) does not produce  giant rings with the exact profile found in the WMAP data.
Despite the fact that we do not know  the exact form of $\Phi_{\text{PIP}}$, we do know  that generically  it is quite different than a typical $\lcdm$ potential: its range is much larger as it varies slowly with distance, and it is invariant under rotation. This is the reason why regardless of the exact form of $\Phi_{\text{PIP}}$, a distinct imprint of a PIP is the existence of giant rings.

When considering the CMB weak lensing signal of a PIP we encounter a similar challenge. We should find an estimator that is as sensitive as possible to the CMB lensing effect of the PIP just by knowing its general properties but without knowing its exact profile. The fact that $\Phi_{\text{PIP}}$ varies slowly with the distance from it implies that, much like in the temperature case of \cite{Kovetz:2010kv}, the CMB lensing signal of the PIP is not located solely in the  small patch of the sky around its location, but rather it is spread  over the entire sky. Hence CMB lensing signal to noise (S/N) calculations of the type done in \cite{Das:2008es, Masina:2009wt, Masina:2010dc, Rathaus:2011xi} that are relevant for anomalously large {\it localized} structures are irrelevant in our case. This also means that SPT and ACT are not useful in facing this challenge, and our best hope to find the PIP's lensing effect is in the Planck data soon to be  available.

\begin{figure*}
\centering
\includegraphics[width=0.7\textwidth]{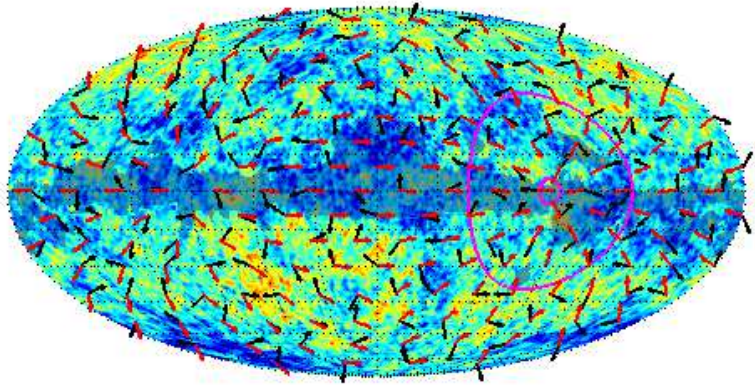}
\caption{A random realization of a CMB temperature map which follows $\lcdm$-statistics and further lensed by a PIP, and with the KQ85 mask suepr-imposed. Even though the effect of WL is unobservable to the naked eye, the direction $\hn_0$ which maximizes the estimator (\ref{eq:tempEstimator}) is in the direction of the PIP, i.e. ($276^\circ, -1^\circ$). Here one can also see the ingredients of the estimator. The magenta circles define the ring $\mathcal{A}_{\hn_0}$ with inner and outer radii $5^\circ$, and $40^\circ$ respectively, the red arrows are the unit vectors $\hat{\mathbf{u}}_i$ pointing from each pixel-center to the direction $\hn_0$, and the black arrows are the unit vectors $\hat{\bm{\nabla} T_i}$ pointing in the direction of the gradient at each pixel.
The normalization factor $\mathcal{N}_{\hn}$ is the number of pixels which lie within the ring $\mathcal{A}_{\hn}$ (and which are outside the mask, in case the mask is applied).
The subtractad average value ($\bar{s}(\hn)$) is not shown in this figure.
}
\label{fig:estimator}
\end{figure*}

Since WL by a spherically symmetric structure deflects light in a radial way, one expects that in the presence of a single anomalous lens,   on average, the radial component of the temperature gradient will be more pronounced than the tangential component. 
This follows from a series expansion of the gradient of the lensed temperature field
\ben
\bm\nabla \tilde{T}(\hn) &\approx& \bm{\nabla} \left[  T(\hn) + \bm{\alpha} \cdot \bm{\nabla}T(\hn) \right] \nonumber \\ 
&=& \bm{\nabla}T(\hn) +\alpha^r \bm{\nabla} \lb\nabla_r T(\hn)\rb \nonumber\\
&& + \lb \nabla_r T(\hn) \rb \bm{\nabla}\alpha^r, \label{eq:estimatorMotivation}
\een
where $\bm{\alpha} = \alpha ~\hat{\mathbf{e}}_r$ is the deflection angle induced by the PIP. The first two terms of 
this series expansion point at random directions, while the last term always points at the radial direction.
This suggests the following estimator
\be \label{eq:tempEstimator2}
\tilde{S}(\hn) = \sum_{\left\{i\left|\hn_i \in \mathcal{A}_{\hn}\right.\right\}}
\frac{\lb  \hat{\bm{\nabla} T_i} \cdot \hat{\mathbf{u}}_i \rb^2} {\mathcal{N}_{\hn}},
\ee
where $\hat{\bm{\nabla} T_i}$ is the unit vector pointing in the direction of the  temperature gradient at the pixel $i$, $\hat{\mathbf{u}}_i$ is the unit vector (on the surface of the 2-sphere) which points from the center of the pixel $i$ in the direction of the point $\hn$ on the sphere,
$\mathcal{A}_{\hn}$ is a ring around the direction $\hn$
and $\mathcal{N}_{\hn}$ is a normalization factor counting the number of pixels around the direction $\hn$ which are within the ring $\mathcal{A}_{\hn}$ (see  Fig.~\ref{fig:estimator}).

In our analysis we take the inner and outer radii of the ring $\mathcal{A}_{\hn}$ to be $5^\circ$, and $40^\circ$, respectively. The reason for the inner radius is to avoid the weak lensing effect of localized large structures like a cosmic void \cite{Das:2008es, Masina:2009wt, Masina:2010dc, Rathaus:2011xi}. The reason for the outer radius is to maximize the effectiveness of our estimator with regard to the detection of a PIP.


This score is still not in a usable form, since
due to discretization  the gradient map is discontinuous. Moreover, it depends on the shape of the pixels. It turns out that this dependence is dominant over the effect we wish to measure. Therefore we must subtract at each point an average value which is characteristic of both the specific point and of the shape of the pixels. We thus correct the estimator (\ref{eq:tempEstimator2}) by subtracting this average value
\be \label{eq:tempEstimator}
S(\hn) = \sum_{\left\{i\left|\hn_i \in \mathcal{A}_{\hn}\right.\right\}}
\frac{\lb  \hat{\bm{\nabla} T_i} \cdot \hat{\mathbf{u}}_i \rb^2} {\mathcal{N}_{\hn}} - \bar{s}(\hn).
\ee
In order to determine the characteristic score at each point $\bar{s}(\hn)$  we generate random  maps and evaluate (\ref{eq:tempEstimator2}). We then average these maps to yield the desired $\bar{s}(\hn)$.

Now that we have an estimator we can check how effective it is in finding a PIP in Planck-like data.
We lens  randomly generated $\lcdm$ maps by  a PIP and see how often the maximum of our estimator points in the direction of the PIP. To  do that we need to know $\Phi_{\text{PIP}}$, which, due to the reasons mentioned above, we do not. Without knowing $\lambda(k)$ it is not feasible to go over all the parameter space associated with $\Phi_{\text{PIP}}$. To be able to proceed we must narrow down this parameter space. 
Despite the fact that  we do not expect (\ref{eq:log}) to be the correct profile, we use it to parameterize $\Phi_{\text{PIP}}$.  The reason is simply that it is easy to work with (\ref{eq:log}) and that our estimator is sensitive only to general properties of $\Phi_{\text{PIP}}$ and not to its exact form. In this parametrization there are two parameters, the PIP's location $r_0$ and $\lambda$.

\begin{figure}
\centering
\includegraphics[width=0.5\textwidth]{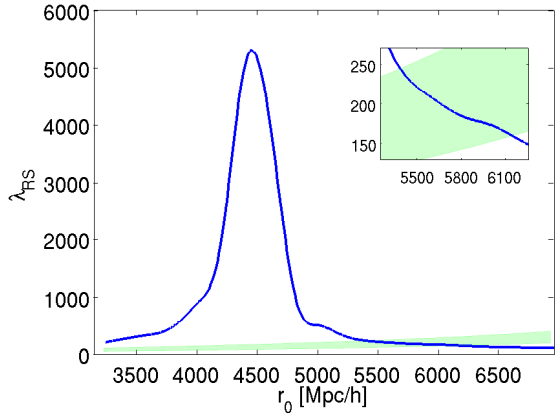}
\caption{The calibration of the coupling $\lambda$ of the PIP so it would explain both the giant rings and the bulk flow anomalies simultaneously. The blue line is the $\lambda$ needed to yield a detection of the rings score with the same significance as reported in \cite{Kovetz:2010kv}, while the green region is the $\lambda$ which would account for a bulk flow of $340 \pm 110 ~\text{km/s}$. The intersection of these, the bulk flow region, zoomed in in the small figure, centered around $\sim 5600~\text{Mpc/h}$  is the region we later use for WL simulations. }
\label{fig:calibration}
\end{figure}

If indeed a PIP is responsible for the giant rings then we can use the results of \cite{Fialkov:2009xm, Kovetz:2010kv}  to estimate the relation between  $\lambda$ and $r_0$. The statistical significance of the giant rings is about $3\sigma$, hence for each $r_0$ we can fix $\lambda^{\text{RS}}(r_0)$ by demanding that the rings score, defined in \cite{Kovetz:2010kv}, identifies correctly the PIP's direction with the same $\sim 3\sigma$ significance. $\lambda^{\text{RS}}(r_0)$ is plotted in Fig. \ref{fig:calibration}. The steep rise in $\lambda^{\text{RS}}(r_0)$ around $4500~\text{Mpc/h}$ is due to the cancellation between the Sachs-Wolfe and integrated Sachs-Wolfe effects induced by the PIP \cite{Fialkov:2009xm}.

\begin{figure}
\centering
\includegraphics[width=0.5\textwidth]{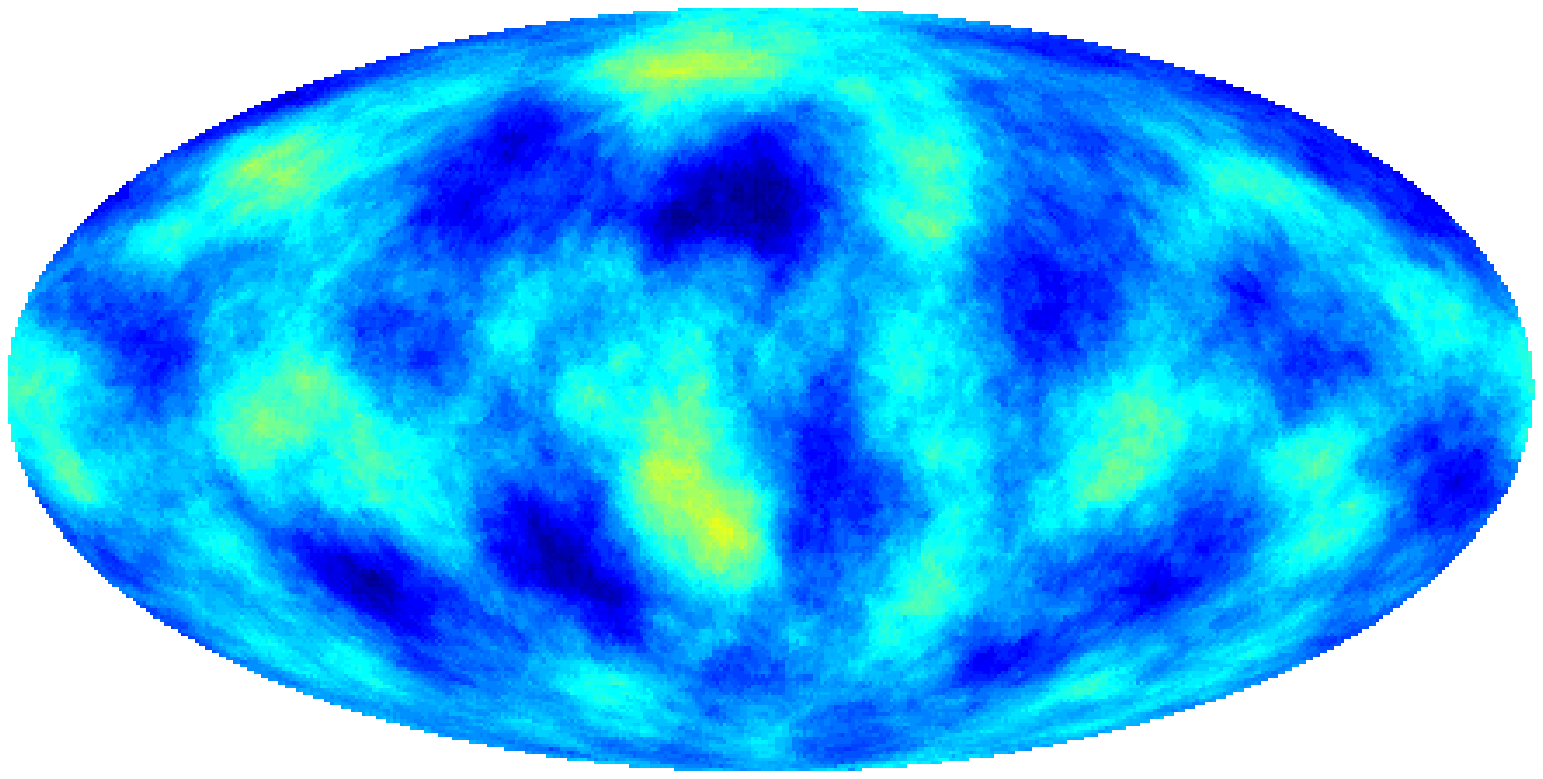}
\includegraphics[width=0.5\textwidth]{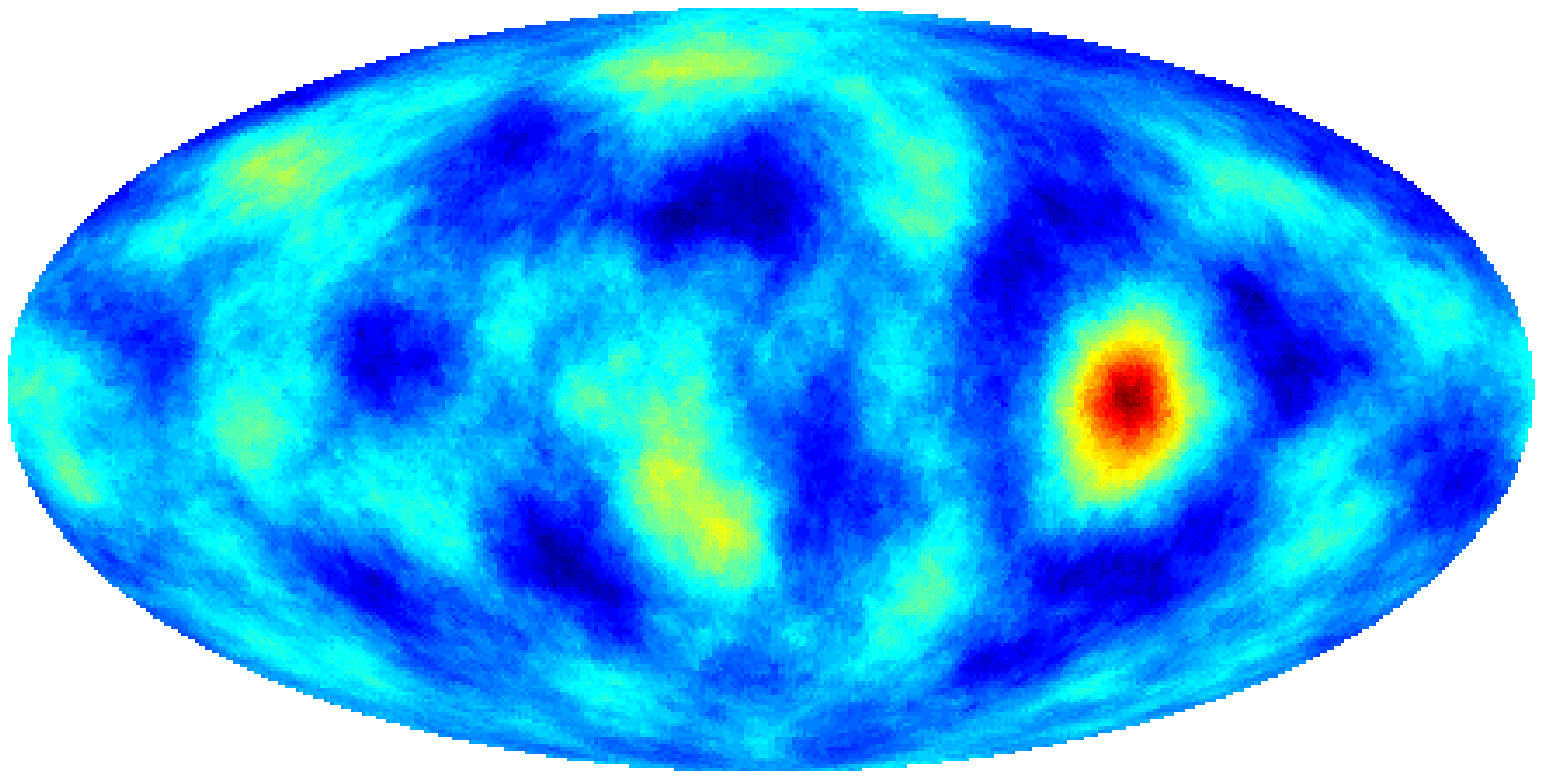}
\includegraphics[width=0.5\textwidth]{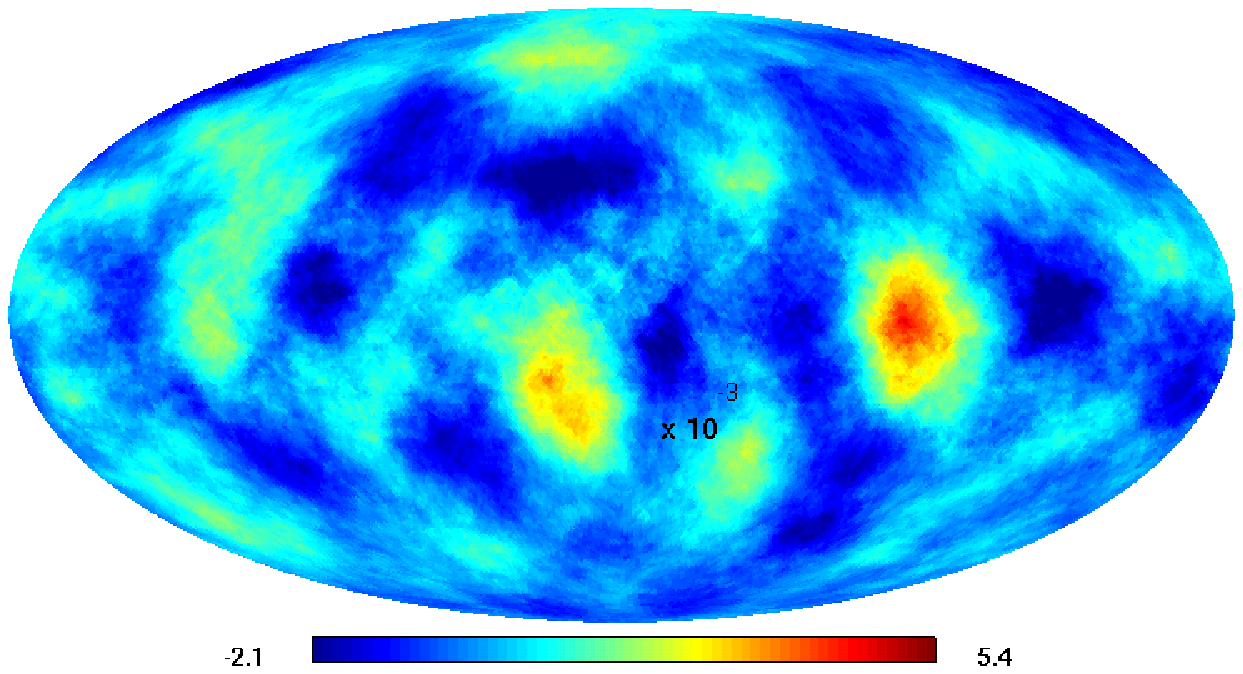}
\caption{An example of the score map ($10^3 S(\hn)$) obtained by the estimator (\ref{eq:tempEstimator}) for a \emph{single} realization of the CMB following $\lcdm$ statistics, and for three cases: (top  to bottom) no PIP at all, PIP at $5600~\text{Mpc/h}$ away from us with $\lambda = 207$ and full sky, and the same PIP with KQ85 mask, all with the same colorbar.
}
\label{fig:example}
\end{figure}

\begin{figure}[t]
\centering
\includegraphics[width=.5\textwidth]{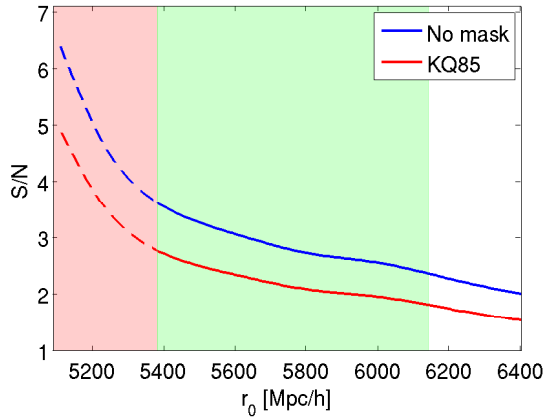}
\caption{Confidence level (in terms of standard deviations) as a function of the distance of the PIP from us and with $\lambda$  calibrated according to Fig. \ref{fig:calibration}, both for the case of no foreground contamination (blue), and for the situation of a masked sky (here KQ85). The red region (dashed lines) is forbidden, as the PIP would induce a bulk flow which exceeds the observed one, and the green region is the flow region. }
\label{fig:sigmaOfR0}
\end{figure}

Due to the above calibration we are left with  a single parameter, $r_0$. The range of $r_0$ is constrained by the bulk flow induced by the PIP. The reason is that the difference between the  bulk flow induced by the PIP and the observed bulk flow cannot be larger than the rms bulk flow in $\lcdm$. There is a lively discussion on the magnitude of the observed bulk flow and whether it is anomalously large or not \cite{Sarkar:2006gh, Feldman:2008rm, arXiv:0809.4041, arXiv:0911.5516, arXiv:1111.0631, Nusser:2011tu, Dai:2011xm}. The range of the observed bulk flow we take is quite wide so it fits with most recent claims, $340 \pm 110~\text{km/s}$. Imposing that the PIP induces a bulk flow in this range defines $\lambda^{\text{BF}}(r_0)$ which is plotted in Fig.~\ref{fig:calibration}.
We see that the region $r_0<5400$ Mpc/h (that corresponds to $z\sim 4.5$) is excluded since $\lambda^{\text{RS}}$ is larger than $\lambda^{\text{BF}}$ which means that the bulk flow induced by a PIP that generates the giant rings is too large to be consistent with observations.

The region $5400<r_0~\text{h/Mpc}<6100$ is of particular interest. A PIP located in this region will induce both the bulk flow and the giant rings which are expected to be nearly aligned (as both point towards the PIP). It is interesting to note that while there is a debate about the magnitude of the bulk flow, as far as we know, all authors agree about the direction. Interestingly enough this direction is nearly aligned with that of the giant rings \cite{Kovetz:2010kv}. We refer to the $5400<r_0~\text{h/Mpc}<6100$ range as the bulk flow range. A PIP located at larger distances, $r_0> 6100$ Mpc/h, that induces the giant rings will induce a smaller bulk flow than the observed one.

Fig. \ref{fig:example} illustrates how the estimator works for a single randomly generated $\lcdm$-like map.
We use $N_{\text{side}} = 1024$  and take $2\leq l \leq 2000$ which is similar to the resolution of Planck.
 In the top panel we show the $S(\hn)$  map (with inner and outer ring radii $5^\circ, 40 ^\circ$ respectively, as discussed above) associated with a  random $\lcdm$ map.
The  $S(\hn)$ map associated with the same random  which is further  lensed by a PIP located at $r_0=5600$ Mpc/h  (in the direction of the giant rings $(276^\circ, -1^\circ)$) is shown in the middle panel with no mask  and in the bottom panel with the KQ85 mask. As expected the full sky map has a sharper peak than the masked sky.

This indicates that the estimator works as it should. What remains to be done is to repeat this simulation many times
and determine the CMB lensing S/N as a function of $r_0$.
Fig. \ref{fig:sigmaOfR0} presents the outcome of this process. The confidence level (in terms of standard deviations) for the detection of a PIP with $\lambda$ calibrated as discussed above via CMB weak lensing is shown as a function of its distance from us.

We see that  in the bulk flow region the CMB lensing is fairly high even with the mask, $\text{S/N} \geq 2$, whereas for larger $r_0$ it drops. We expect  this  to be true regardless of the exact profile $\Phi_{\text{PIP}}(r)$. This implies that if a PIP is responsible both for the giant rings and for the bulk flow then we should be able to further test the model with  CMB weak lensing. However if the giant rings are generated by the PIP and the bulk flow is not, then it is unlikely that its CMB lensing signal is detectable.

\vspace{10mm}

\noindent {\bf Acknowledgements}


We thank A. Ben-David, A. Fialkov and E. Kovetz for discussions.
We acknowledge the use of the Legacy Archive
for Microwave Background Data Analysis (LAMBDA)
\cite{lambda} and the use of the HEALPix package \cite{healpix}. This work
is supported in part by the Israel Science Foundation
(grant number 1362/08) and by the European Research
Council (grant number 203247).

\end{document}